\documentstyle[eqsecnum,aps]{revtex}

\begin{document}
\title{The imposition of Cauchy data to the Teukolsky 
equation I: The nonrotating case}
\author{Manuela CAMPANELLI\thanks{
Electronic address: manuela@mail.physics.utah.edu}}
\address{Department of Physics, University of Utah\\
201 JBF, Salt Lake City, UT 84112, USA}
\author{and Carlos O. LOUSTO\thanks{%
Electronic address: lousto@iafe.uba.ar}}
\address{Instituto de Astronom\'\i a y F\'\i sica del Espacio,\\
Casilla de Correo 67, Sucursal 28\\
(1428) Buenos Aires, Argentina}
\date{\today }
\maketitle

\begin{abstract}
Gravitational perturbations about a Kerr black hole in the Newman-Penrose
formalism are concisely described by the Teukolsky equation. New numerical
methods for studying the evolution of such perturbations require not only
the construction of appropriate initial data to describe the collision of
two orbiting black holes, but also to know how such new data must be imposed
into the Teukolsky equation.
In this paper we show how Cauchy data can be incorporated explicitly into
the Teukolsky equation for non-rotating black holes. The Teukolsky function $%
\Psi $ and its first time derivative $\partial_t \Psi $ can be
written in terms of
only the $3$-geometry and the extrinsic curvature in a gauge invariant way.
Taking a Laplace transform of the Teukolsky equation incorporates initial
data as a source term. We show that for astrophysical data the
straightforward Green function method leads to divergent integrals that can
be regularized like for the case of a source generated by a particle
coming from infinity.
\end{abstract}



\section{INTRODUCTION AND OVERVIEW}

Black holes coalescence is considered to be one of the strongest
astrophysical sources of gravitational radiation and a primary candidate to
be detected by gravitational wave observatories now under construction. For
this reason, several groups of researchers are now attacking the problem of
solving Einstein's equations numerically, with supercomputers \cite
{GrandChallenge}. Many difficulties must still be solved in this approach,
such as the presence of instabilities in the numerical evolution due to the
nonlinearity of the Einstein's equations and that of finding a better
qualitative prescription for truly astrophysical initial data representing
black holes collisions.

Meanwhile, the perturbation theory represents a valid and complementary
approach to full numerical simulations, with some clear advantages with
respect the supercomputer project: It is a considerably more economical
approach and it is semianalytical. This latter aspect is a very important
one since it allows to study some fundamental and conceptual problems. Among
the new theoretical results in perturbation theory stands out the close
limit approximation\cite{PP94} which approximates the collision of two black
holes by a single perturbed one. The subsequent extension to second order
perturbations\cite{GNPP96}, and to moving (head-on) holes\cite{BAABPPS97},
confirmed the success of this approximation. Also recently\cite
{LP97a,LP97b,LP98}, the collision of a small and a big black hole by
perturbative methods have been revisited in order to incorporate
non-vanishing initial data. However, only perturbations about Schwarzschild
black holes have been studied so far, by use of the Zerilli-Moncrief\cite
{Z70,M74} equation (a gauge invariant description). The success of these
techniques encourages now to extend them to the more realistic rotating
background. In particular, for the most plausible astrophysical scenario of
a final single rotating hole, the first order ``curvature'' perturbations
are compactly described by the Teukolsky equation\cite{T72,T73}, which in
the non-rotating case reduces to the so called Bardeen-Press equation\cite
{BP73}.

The Teukolsky equation has been studied already in the early seventies in
the frequency domain. In order to avoid the important problem of the
imposition of the initial data into that equation, the computation of the
gravitational radiation have been carried out in the case of unbounded
particle trajectories (or circular motion)\cite{D79}. The divergent
integrals encountered when one wants to calculate the gravitational
radiation generated by a test mass falling into a black hole from infinity
using the standard Green function, have been made finite by systematically
discarding infinite surface terms\cite{DS79}. In a different way to tackle
this problem, Nakamura and Sasaki \cite{SN82} transformed, in the frequency
domain, the Teukolsky equation into other ``more regular'' equation. Only
very recently, Poisson\cite{P97} (in the non-rotating case) and Campanelli
and Lousto\cite{CL97} (for the rotating hole) showed that there is nothing
intrinsically wrong with the radial Teukolsky equation when dealing with
unbounded source terms, and that the divergent integrals can be regularized
in a natural way. In the Appendix, we will show how those results about
regularization also apply when dealing with initial data for a {\it bounded}
problem (like for the close limit case).

The problem of imposition of initial data in the Teukolsky equation is not,
of course, restricted to the frequency domain form of the equation. Very
recently an evolution code to integrate the Teukolsky equation in its
time-domain form have been developed \cite{KLPA97}. However, until now the
evolution of initial data have been restricted to the simple case of a
bell-shaped burst, which cannot represent accurately realistic astrophysical
initial data for the late stages of binary black hole coalescence. The
imposition of the initial data into the Teukolsky equation, to our
knowledge, seems not to have a clear framework yet. Consequently, in this
paper we address to the important issue of the imposition of the initial
data into the Teukolsky equation. In the non-rotating case, the gauge
invariant Teukolsky functions $\psi _0$ (ingoing) and $\psi _4$ (outgoing),
and their first time derivatives can be completely expressed in terms of the
Moncrief waveform (and its time derivative), and then, in terms of the $3$%
-geometry and the extrinsic curvature. By evaluating these functions on a
constant time hypersurface for any specific astrophysical initial data, one
is in principle able to numerically evolve the Teukolsky equation and to
study very interesting astrophysical scenarios, like the close limit
approximation\cite{CKL98}. The imposition of initial data into the
Bardeen-Press equation\cite{BP73} can suggest what features need to be
generalized in the rotating case.

The paper is organized as follows: Section\ II is devoted to the problem
of expressing the Teukolsky functions $\psi _0$ and $\psi _4$ on a given
hypersurface in terms of the $3$-geometry and the extrinsic curvature of
that hypersurface. The computations are specifically performed in the
Regge-Wheeler gauge\cite{RW57} for both even and odd parity waves and then
reexpressed in a gauge invariant way by use of the Moncrief functions. In
Sec.\ III we discuss the applicability of these relations and the
possibility of generalization of these results to the
rotating case. In the Appendix we briefly review the
Teukolsky equation in its frequency domain and use the Laplace
transform to mathematically incorporate initial data into the radial
equation as an effective source term.
We explicitly replace Brill-Lindquist initial data into the
source term of the radial Teukolsky equation. The source term is shown to
lead to a divergent integral as the domain of integration in the coordinate
$r$ extends to infinity. The regularization procedure already used for
unbounded sources is employed successfully in the present case as
described in the Appendix.

The notation and conventions employed in this paper are those of Misner,
Thorne and Wheeler \cite{MTW73}. That is, the metric signature is $(-~+~+~+)$;
 the Riemann tensor is defined by the equation $R_{~\beta \gamma \delta
}^\alpha =\Gamma _{~\beta \gamma ,\delta }^\alpha -\Gamma _{~\beta \delta,
\gamma }^\alpha +\Gamma _{~\beta \gamma }^\sigma \Gamma _{~\delta \sigma
}^\alpha -\Gamma _{~\beta \delta }^\sigma \Gamma _{~\gamma \sigma }^\alpha $
and the Ricci tensor by equation $R_{\mu \nu }=R_{~\mu \alpha \nu }^\alpha $.
 Throughout the paper an overdot indicates differentiation with respect to
the time variable $t$, a prime differentiation with respect to the radial
coordinate $r$, an overbar represents complex conjugation, and units are
such that $G=c=1$. The conventions for the NP quantities, i. e. spin
coefficients, operators (denoted by an overhat and the Kinnersley tetrad),
employed in this paper are those given in the table of Ref. \cite{CL97}.

\section{INITIAL DATA FOR THE TEUKOLSKY EQUATION}

In the Newman--Penrose formalism, gravitational perturbations of the Kerr
metric can be decoupled and described by the Teukolsky\cite{T73} equation.
In the Boyer-Lindquist coordinates $(t,r,\theta ,\varphi )$ this equation
takes the following form 
\begin{eqnarray}
&&\ \ \Biggr\{\left[ a^2\sin ^2\theta -\frac{(r^2+a^2)^2}\Delta \right]
\partial _{tt}-\frac{4Mar}\Delta \partial _{t\varphi }-2s\left[ (r+ia\cos
\theta )-\frac{M(r^2-a^2)}\Delta \right] \partial _t  \nonumber \\
&&\ \ +\,\Delta ^{-s}\partial _r\left( \Delta ^{s+1}\partial _r\right) +%
\frac 1{\sin \theta }\partial _\theta \left( \sin \theta \partial _\theta
\right) +\left[ \frac 1{\sin ^2\theta }-\frac{a^2}\Delta \right] \partial
_{\varphi \varphi }  \label{master} \\
\ &&+\,2s\left[ \frac{a(r-M)}\Delta +\frac{i\cos \theta }{\sin ^2\theta }%
\right] \partial _\varphi -s\left( s\cot ^2\theta -1\right) \Biggr\}\Psi
=4\pi \Sigma T,  \nonumber
\end{eqnarray}
where $M$ is the mass of the black hole, $a$ its angular momentum per unit
mass, $\Sigma \equiv r^2+a^2\cos ^2\theta $, $\Delta \equiv r^2-2Mr+a^2$,
and $s$ is the spin parameter characterizing the perturbation.

The field (where $\rho =1/(r-ia\cos \theta )$ ) 
\begin{equation}  \label{psi}
\Psi (t,r,\theta ,\varphi )=\left\{ 
\begin{array}{ll}
\rho ^{-4}\psi _4^{(1)}\equiv -\rho ^{-4}C_{n\bar mn\bar m} & {\rm for}~~s=-2
\\ 
\psi _0^{(1)}\equiv -C_{lmlm} & {\rm for}~~s=+2~
\end{array}
\right. ,  \label{RH}
\end{equation}
represents either the outgoing radiative part of the perturbed Weyl tensor, $%
\psi _4^{(1)}$ ($s=-2$), or the ingoing radiative part, $\psi _0^{(1)}$ ($%
s=+2$). The corresponding unperturbed quantities $\psi _4^{(0)}$ and $\psi
_0^{(0)}$ vanish identically in the Kerr spacetime. The knowledge of only
one of these two first order complex NP components, formed by projecting the
perturbed Weyl tensor along the Kinnersley tetrad, is sufficient\cite{W72}
to uniquely and completely specify the gravitational perturbation (up to
changes of the Kerr parameters $a$ and $M$). In fact, $\psi _4^{(1)}$ and $%
\psi _0^{(1)}$ are gauge invariant under both coordinate transformations and
infinitesimal tetrad (Lorentz) rotations. These two scalars carry
information, in their real and imaginary parts, about the two dynamical
degrees of freedom of the perturbed field. More precisely, in the
nonrotating case, the real part of
$\Psi $ contains information about the even parity (or polar) perturbations
of the metric (which are invariant under the transformation $\varphi
\leftrightarrow -\varphi $) while its imaginary part contains information of
odd parity (or axial) perturbations of the metric (that change sign under a
sign reversal of $\varphi $). The source term $T$ in Eq.\ (\ref{master}) is
constructed out of the $T_{\mu \nu }^{(1)}$ and its expression can be found
in Refs. \cite{T73,CL97}.

In order to study perturbations around a Kerr hole one has to supplement the
Teukolsky equation with initial data that allow to start the
integration. So far the Teukolsky equation has been studied (in the 
frequency domain) for perturbations generated by particles coming from
infinity, i.e. with vanishing initial data.
There are, though,  astrophysically relevant situations where
initial data contribution are important. This is the case of the very
successful close limit approximation\cite{PP94}.
We can divide the initial value problem for the Teukolsky equation into
two subproblems: First, one must find {\it what} Cauchy
data can better represent astrophysical scenarios for colliding black holes.
Slices of constant Boyer-Linquist time of the Kerr geometry give a
non-conformally flat three-geometry, making it incompatible with all
initial data studied so far.
It is not known wheter a slice of the Kerr geometry exists such
that its three-geometry is conformally flat. It seems that, on the other
hand, it is simpler to find solutions to the initial value problem compatible
with Kerr perturbations on the usual hypersurfaces\cite{BIMW98,BP98}.
In the present paper we will be more concerned with the question of {\it how}
to impose initial data to the Teukolsky equation.

Cauchy data in General Relativity consists of $(g_{ik},K_{ik})$ and a
three-dimensional spacelike hypersurface $\Sigma _t$, where $g_{ik}$ is a
Riemannian three-metric on $\Sigma _t$, and\cite{MTW73} 
\begin{equation}
K_{ik}={\frac 1{2N}}\left( N_{i,k}+N_{k,i}-\partial _tg_{ik}-2\Gamma
_{ik}^pN_p\right)  \label{curvextr}
\end{equation}
is the extrinsic curvature which describes the ``embedding'' of $\Sigma _t$
in the spacetime, written in terms of the lapse $N=(-g^{tt})^{-1/2}$ and of
the shift $N_i=g_{ti}$.
In perturbation theory, to first order in the metric perturbation about the
background metric $g_{\mu \nu }^{(0)}$, the Cauchy data for Einstein's
equations, $\delta G_{\mu \nu }^{(1)}=8\pi \delta T_{\mu \nu }^{(1)}$, are
given in terms of the first order perturbations of the metric, $%
h_{ik}^{(1)}$, and in terms of the first order extrinsic curvature, $%
K_{ik}^{(1)}$, on a constant time hypersurface. In the case of the Teukolsky
equation, the first order initial value problem must be translated into
terms of $\Psi |_{t=0}$ and $\partial_t \Psi |_{t=0}$. As we will show in this
Section relating $(h_{ik}^{(1)},K_{ik}^{(1)})$ at the initial hypersurface $%
t=0$ with $(\Psi |_{t=0},\partial_t \Psi |_{t=0})$ is not straightforward.

The starting point is given by Chrzanowski \cite{C75}, who related the
Teukolsky functions to metric perturbations 
\begin{eqnarray}
\psi _4^{(1)} &=&\frac 12\Biggr\{(\bar \delta -3\alpha -
\bar \beta-\pi +\bar \tau)(\bar \delta -2\alpha -2\bar \beta +\pi 
+\bar \tau )h_{nn}+(\hat \triangle -\mu -\bar \mu -3\gamma +\bar \gamma )
(\hat \triangle +\mu -\bar \mu -2\gamma +2\bar \gamma )
h_{\bar m\bar m}  \nonumber \\
&&\ \ \ \ -2\left[ (\hat \triangle -\mu -\bar \mu -3\gamma +\bar
\gamma )(\bar \tau +\pi )+(\bar\delta -3\alpha -\bar\beta -\pi+\bar\tau )(%
\hat \triangle -2\bar \mu -2\gamma )\right] h_{n\bar m}\Biggr\}  \label{psi4}
\end{eqnarray}
and 
\begin{eqnarray}
\psi _0^{(1)} &=&\frac 12\Biggr\{(\hat \delta +\bar \alpha 
+3\beta-\bar\pi+\tau )(\hat \delta +2\alpha +2\beta -%
\bar \pi-\tau )h_{ll}+(\hat D+\rho+\bar \rho+3\epsilon-\bar\epsilon )
(\hat D+\bar \rho-\rho+2\epsilon-2\bar\epsilon )h_{mm}  \nonumber
\\
&&\ \ \ \ -2\left[ (\hat D+\rho+\bar \rho +3\epsilon-\bar\epsilon)
(\tau +\bar\pi )+(\hat \delta +\bar \alpha +3\beta-\bar\pi+\tau )
(\hat D+\bar \rho +2\epsilon)\right] h_{lm}\Biggr\}  \label{psi0}
\end{eqnarray}
where $h_{nn}=n^\mu n^\nu h_{\mu \nu }$, $h_{lm}=l^\mu m^\nu h_{\mu \nu }$,
etc, contain not only three-metric perturbations but all the components of
the first order metric perturbations (here $l,n,m$ and $\stackrel{\_}{m}$
are the null vectors of the Kinnersley tetrad). This represents a practical
problem at the moment of imposing initial data, since one is normally given
only the 3-geometry and the extrinsic curvature, not the 4-geometry
and its time derivative (nor we should need it). The point
is that $\psi _4$ (and $\psi _0$ ) are gauge invariant quantities, but we do
not have them explicitly expressed in terms of purely hypersurface
quantities.

Many simplifications in the analysis are possible when the background has
spherical symmetry. In the Schwarzschild black hole case expressions \ (\ref
{psi4})-(\ref{psi0}) reduce to 
\begin{eqnarray}
\psi _4^{(1)} &=&\frac 1{16}\Biggr\{\frac 1{r^2}\left( \partial _\theta
-\cot \theta -\frac i{\sin \theta }\partial _\varphi \right) \left(
\partial _\theta -\frac i{\sin \theta }\partial _\varphi
\right)\left(h_{tt}-2h_{rt}f+h_{rr}f^2\right)   \nonumber \\
&&\ \ \ \ +\left( \partial _t-f\partial _r+f^{\prime }-\frac {2f}{r}\right)
\left( \partial _t-f\partial _r\right) \frac 1{r^2}\left( h_{\theta
\theta }-\frac{h_{\varphi \varphi }}{\sin ^2\theta }-2i\frac{h_{\theta
\varphi }}{\sin \theta }\right)   \label{psis4} \\
&&\ +\frac 2{r^2}\left( \partial _t-f\partial _r+f^{\prime }\right) 
\left( \partial _\theta -\cot \theta -\frac i{\sin \theta }
\partial _\varphi \right)\left[h_{t\theta }-i\frac{h_{t\varphi }}
{\sin \theta }-f\left( h_{r\theta }-i\frac{%
h_{r\varphi }}{\sin \theta }\right) \right] \Biggr\}  \nonumber
\end{eqnarray}
and 
\begin{eqnarray}
\psi _0^{(1)} &=&\frac 1{4}\Biggr\{\frac 1{r^2} 
\left( \partial _\theta -\cot \theta +\frac i{\sin \theta }
\partial _\varphi \right) \left( \partial _\theta +%
\frac i{\sin \theta }\partial _\varphi\right)
\left(h_{tt}f^{-2}-2h_{rt}f^{-1}+h_{rr}\right)   \nonumber \\
&&+\left( f^{-1}\partial _t+\partial _r+\frac 2r\right)
\left(f^{-1}\partial _t+\partial _r\right)
\frac 1{r^2}\left(h_{\theta \theta }-
\frac{h_{\varphi \varphi }}{\sin ^2\theta }+2i\frac{%
h_{\theta \varphi }}{\sin \theta }\right)   \label{psis0} \\
&&-\frac 2{r^2}\left( f^{-1}\partial _t+\partial _r\right) 
\left( \partial _\theta -\cot \theta +\frac i{\sin \theta }
\partial _\varphi \right)\left[ f^{-1}\left(h_{t\theta }
+i\frac{h_{t\varphi }}{\sin \theta }\right) +h_{r\theta }+i\frac{%
h_{r\varphi }}{\sin \theta }\right] \Biggr\}  \nonumber
\end{eqnarray}
where $f=1-2M/r$ and $f^{\prime }=2M/r^2$ .

The imposition of spherical symmetry carry also the following computational
(but not necessarily crucial) advantage: the multipole decomposition of the
metric perturbations in terms of spin-weighted
harmonics $_{-2}Y_\ell^m(\theta)$ can be
performed\cite{RW57,M74}, and even and odd parity perturbations decouple so
they can be considered independently. Below we shall decompose all metric
perturbations in multipoles with (implicit) index $\ell m$ (not to be
confused with the tetrad vectors).

\subsection{Even or polar parity waves}

{}From now on we will consider only $\psi _4$ because of its better behavior
for numerical integration compared to $\psi _0$ and because the radiated
energy can be described in terms of only $\psi _4$. Similar equations will
hold for $\psi _0$ . Even parity perturbations are given by the
real part ($\text{Re} $) of $\psi _4$. In the Regge-Wheeler (RW)
gauge, $h_0=h_1=G=0 $ ($h_{t\theta }=h_{t\varphi }=h_{r\theta }=h_{r\varphi
}=0,h_{\theta \theta }=h_{\varphi \varphi }/\sin ^2\theta $), we have $h_{m%
\bar m}=h_{nm}=h_{mm}=h_{lm}=0$, then Eq. (\ref{psis4}) and its time
derivative are given by 
\begin{eqnarray}
\text{Re}\, 
\psi _4^{(1)} &=&-\frac{\left( 1-\frac{2M}r\right) }{16r^2}\sum_{\ell m}%
\sqrt{(\ell -1)\ell (\ell +1)(\ell +2)}(H_0-2H_1+H_2)\;_{-2}Y_\ell^m
\label{RWSpsi4} \\
\text{Re}\, 
\partial _t\psi _4^{(1)} &=&-\frac{\left( 1-\frac{2M}r\right) }{16r^2}%
\sum_{\ell m}\sqrt{(\ell -1)\ell (\ell +1)(\ell +2)}(\partial
_tH_0-2\partial _tH_1+\partial _tH_2)\;_{-2}Y_\ell^m
\end{eqnarray}
Using Einstein's equations in the RW gauge, Zerilli's\cite{Z70}, Eq.
(C7e) gives, 
\begin{equation}
H_0=H_2+\frac{32\pi r^2}{\sqrt{2(\ell -1)\ell (\ell +1)(\ell +2)}}F_{\ell
m}~,  \label{RWEvenEinstein}
\end{equation}
where $F_{\ell m}$is a source term given in Table III of Ref.\cite{Z70}. The
above expressions allow us to write Eqs. (\ref{RWSpsi4}) in terms of $H_2$ ,$%
H_1$, and its time derivatives [Note that here and below there is an
implicit index $\ell m$ coming from the multipole decomposition of the
metric perturbations.] The following step is to restore the gauge
invariance, lost in principle when, for the sake of simplicity, we
have chosen to work in the Regge-Wheeler gauge. To do so, we will use a
result of Ref. \cite{L98} where the metric perturbations in the RW gauge
have been expressed in terms of $\phi _M$ , the Moncrief invariant function 
\begin{equation}
\phi _M=\frac r{\lambda +1}\left[ K+\frac{r-2M}{\lambda r+3M}\left(
H_2-r\partial _rK\right) \right] +\frac{\left( r-2M\right) }{\lambda r+3M}%
\left( r^2\partial _rG-2h_1\right)\,;\quad
\lambda=\frac12(\ell-1)(\ell+2),  \label{psimon}
\end{equation}
which is expressed entirely in terms of the 3-geometry and is gauge
invariant under changes of coordinates in the spacetime. The other piece of
initial data, $\partial _t\phi _M$ , is expressed in terms of the extrinsic
curvature $K_{ij}$ as in Ref. \cite{AP96a}

\begin{eqnarray}
\partial _t\phi _M &=&\frac{-2r}{\lambda +1}\left[ \sqrt{1-\frac{2M}r}%
K_{(K)}+\frac{r-2M}{\lambda r+3M}\left( \left( 1-\frac{2M}r\right)
^{3/2}K_{rr}-r\partial _r\left( \sqrt{1-\frac{2M}r}K_{(K)}\right) \right)
\right]  \nonumber \\
&&\ +\frac{\left( r-2M\right) }{\lambda r+3M}\left( r^2\partial _r\left( 
\sqrt{1-\frac{2M}r}K_{(G)}\right) -2\sqrt{1-\frac{2M}r}K_{r\theta }^{\text{%
even}}\right)  \label{psimonpunto}
\end{eqnarray}
where $K_{(K)}$ and $K_{(G)}$ are respectively the ''$K$ ''and ''$G$ ''parts
of the extrinsic curvature, in analogy to the Regge-Wheeler\cite{RW57}
decomposition of the metric tensor (see also a similar decomposition for $%
\widehat{K}$ in Eqs. (2.32)-(2.35) of Ref. \cite{LP97b}). Also, here and
below, extrinsic curvature components (of even and odd parity) have indices $%
\ell m$ implicit.

In Ref. \cite{L98} it was found by use of the definition (\ref{psimon}) and
the Hamiltonian constraint (see for instance Zerilli's equation (C7a) in Ref.%
\cite{Z70}) 
\begin{eqnarray}
H_2=-\frac{9M^3+9\lambda M^2r+3\lambda ^2Mr^2+\lambda ^2(\lambda +1)r^3}{%
r^2(\lambda r+3M)^2}\,\phi _M+\frac{3M^2-\lambda Mr+\lambda r^2}{r(\lambda
r+3M)}\phi _M^{\prime }+(r-2M)\phi _M^{\prime \prime }  \label{H2}
\end{eqnarray}
here and below, primes denote $\partial _r$ .

{}From the momentum constraint, equation (C7b) in Ref.\cite{Z70}

\begin{equation}
H_1=r\partial _t\phi _M^{\prime }+\frac{\lambda r^2-3\lambda Mr-3M^2}{%
(r-2M)(\lambda r+3M)}\partial _t\phi _M.  \label{H1}
\end{equation}

The time derivatives of these metric coefficients in terms of $\phi _M$ have
also been found in Ref. \cite{L98}

\begin{eqnarray}
\partial _tH_2 &=&(r-2M)\partial _t\phi _M^{\prime \prime }+\frac{%
3M^2-\lambda Mr+\lambda r^2}{(\lambda r+3M)r}\partial _t\phi _M^{\prime } 
\nonumber \\
&&-\frac{9M^3+9\lambda M^2r+3\lambda ^2Mr^2+\lambda ^2(\lambda +1)r^3}{%
(\lambda r+3M)^2r^2}\partial _t\phi _M  \label{H2punto}
\end{eqnarray}
and from the time derivative of Eq. (\ref{H1}), $\partial _tH_1=r\partial
_t^2\phi _M^{\prime }+(\lambda r^2-3\lambda Mr-3M^2)/(r-2M)/(\lambda
r+3M)\partial _t^2\phi _M$ , and the Zerilli\cite{Z70} wave equation in
vacuum, $\partial _t^2\phi _M=\partial _{r^{*}}^2\phi _M-V_\ell (r)\phi _M$,
we obtain

\begin{eqnarray}
\partial _tH_1 &=&\frac{(r-2M)^2}r\phi _M^{\prime \prime \prime }+\frac{%
(r-2M)\left( \lambda r^2+3\lambda Mr+15M^2\right) }{r^2(\lambda r+3M)}\phi
_M^{\prime \prime }  \nonumber \\
&&+\left\{ \frac{2M\left[ -\lambda r^2+3\left( \lambda -2\right)
Mr+15M^2\right] }{r^3(\lambda r+3M)}-rV_\ell (r)\right\} \phi _M^{\prime }
\label{H1punto} \\
&&-\left\{ \frac{\lambda r^2-3\lambda Mr-3M^2}{(r-2M)(\lambda r+3M)}V_\ell
(r)+rV_\ell ^{\prime }(r)\right\} \phi _M  \nonumber
\end{eqnarray}
where $V_\ell (r)$ is the Zerilli potential\cite{Z70}. Here it is evident
that in order to rewrite $\psi _4$ and $\partial _t\psi _4$ in terms of
hypersurface data we had to use the evolution equations, not only the
constraints.

For the sake of simplicity we have written Eqs. (\ref{H2})-(\ref{H1punto})
in the sourceless case, but source terms can be straightforwardly included 
\cite{L98}. The above relations together with (\ref{RWEvenEinstein}) provide
the expressions that allow us to write $\psi _4$ and $\partial _t\psi _4$ in
terms only of $\phi _M$ and $\partial _t\phi _M$ (and radial derivatives of
them). This relation holds for all $r$ and $t$ and is gauge invariant since
all involved final expressions are (in spite of having used the RW gauge as
an intermediate step), in particular, this relation holds at the
hypersurface where we want to give initial data. Since $\phi _M$ can be
written only in terms of the perturbed 3-geometry and $\partial _t\phi _M$
in terms of only the extrinsic curvature (cfr. Eqs. (\ref{psimon}) and (\ref
{psimonpunto})) , we will have, at $\Sigma _t$

\begin{equation}
\psi _4=\psi _4(h_{ij},K_{ij}),\;\;\partial _t\psi _4=\partial _t\psi
_4(h_{ij},K_{ij})  \label{final}
\end{equation}
our desired relations. Note that, in general, even for stationary, {\it %
time-symmetric} initial data neither $\psi _4$ nor $\partial _t\psi _4$
vanish. This is in contrast with what happens with the corresponding initial
data for the Moncrief wave form, where $\partial _t\phi _M=0,$ and with the
initial data taken in Ref. \cite{KLPA97}.

\subsection{Odd or axial parity waves}

For odd parity perturbations we shall parallel the procedure we just
followed in the even parity case. Let us recall that odd parity is described
by the imaginary part ($\text{Im} $) of
$\psi _4$ . In the RW gauge ($h_2=0$) we have 
\begin{equation}
\text{Im}\, \psi _4^{(1)}=-\frac 1{8r^2}\left[ \partial _t-\left( 1-\frac{2M}r%
\right) \partial _r+\frac{2M}{r^2}\right] \sum_{\ell m}\sqrt{(\ell -1)\ell
(\ell +1)(\ell +2)}\left( h_0-\left( 1-\frac{2M}r\right) h_1\right)
\;_{-2}Y_\ell^m  \label{psiodd}
\end{equation}
and 
\begin{equation}
\text{Im}\, 
\partial _t\psi _4^{(1)}=-\frac 1{8r^2}\left[ \partial _t-\left( 1-\frac{%
2M}r\right) \partial _r+\frac{2M}{r^2}\right] \sum_{\ell m}\sqrt{(\ell
-1)\ell (\ell +1)(\ell +2)}\left( \partial _th_0-\left( 1-\frac{2M}r\right)
\partial _th_1\right) \;_{-2}Y_\ell^m  \label{psioddpunto}
\end{equation}
here $h_1$ and $h_0$ are the odd parity metric perturbations (not to confuse
with the same symbols used for the even parity perturbations, that in the RW
gauge are taken to vanish).

The corresponding gauge invariant waveform for odd parity given by Moncrief%
\cite{M74} is 
\begin{equation}
Q=\frac 1r\left( 1-\frac{2M}r\right) \left[ h_1+\frac 12\left( \partial
_rh_2-\frac 2rh_2\right) \right]  \label{Q}
\end{equation}
and 
\begin{equation}
\partial _tQ=\frac 1r\left( 1-\frac{2M}r\right) \left[ \sqrt{1-\frac{2M}r}%
K_{r\varphi }^{\text{odd}}+\frac 12\left( \partial _r\left( \sqrt{1-\frac{2M}%
r}K_{\theta \theta }^{\text{odd}}\right) -\frac 2r\sqrt{1-\frac{2M}r}%
K_{\theta \theta }^{\text{odd}}\right) \right] .  \label{Qpunto}
\end{equation}

Again, in the RW gauge ($h_2=0$) we can write (from Eq. (\ref{Q})) 
\begin{equation}
h_1=\frac r{1-\frac{2M}r}Q,\;\;\partial _th_1=\frac r{1-\frac{2M}r}\partial
_tQ,  \label{h1}
\end{equation}
{}from Zerilli's$\cite{Z70}$ Eq. (C6c), one obtains 
\begin{equation}
\partial _th_0=\left( 1-\frac{2M}r\right) \left[ \partial _r\left( rQ\right)
+\frac{8\pi ir^2D_{\ell m}}{\sqrt{(\ell -1)\ell (\ell +1)(\ell +2)/2}}\right]
\label{h0punto}
\end{equation}
where $D_{\ell m}$ is a source term given in Table III of Ref. \cite{Z70},
and finally $h_0$ can be obtained from the definition of extrinsic
curvature, i.e. Eq. (\ref{curvextr}), $\partial _th_2=2\left( \sqrt{1-\frac{%
2M}r}K_{\varphi \varphi }-h_0\right) $ , then in the RW gauge 
\begin{equation}
h_0=\sqrt{1-\frac{2M}r}K_{\varphi \varphi }  \label{h0}
\end{equation}
In spite of having constructed Eqs. (\ref{h1})-(\ref{h0}) in the RW gauge,
when we replace their right hand side in the expressions for
$\text{Im}\, \psi _4$and $\text{Im}\,  \partial _t\psi _4$,
given by the Eqs. (\ref{psiodd}) and (\ref{psioddpunto}),
they will be gauge invariant if we consider the general form
of $Q$ and $K_{\varphi \varphi }$ . We have then succeeded in our objective
of expressing also the odd parity perturbations (the imaginary part of $\psi
_4$ ) in terms only of the 3-geometry and the extrinsic curvature.

\section{DISCUSSION}

In this paper we have faced (and solved for the $a=0$ case) the problem of
giving initial data to the equation that describe in a gauge invariant way
perturbations around the Kerr geometry, i.e. the Teukolsky equation in the
time and frequency domains. We first observed that the connection between
metric perturbations and gauge invariant objects that fully describe
gravitational perturbations $\psi _4$ (or $\psi _0)$ as given by Chrzanowski%
\cite{C75}, is not explicitly written only in terms of the 3-geometry and
the extrinsic curvature. This poses a difficulty when one wants to use this
relation to impose initial data, since, normally one is given only the
3-geometry and the extrinsic curvature on a given hypersurface (and that
should be all we need). In the nonrotating ( $a=0$ ) case (Schwarzschild
background), we have been able to reexpress $\psi _4$ (and $\psi _0)$ and
its time derivative in terms only of the 3-geometry and the extrinsic
curvature. To do so, we have momentarily gone to the Regge-Wheeler gauge and
then restored gauge invariance by writing the metric perturbations in terms
of $\phi _M$ , the Moncrief's invariant waveform and its time derivative
(and radial derivatives on the hypersurface). As a by-product we obtained 
relations between the waveforms of Teukolsky and Moncrief (of course, they
are not independent of each other) on any given hypersurface.
These relations (for both the even and odd parity cases) represent the
generalization of the Chandrasekhar transformations
in the time domain when source terms are present. It is worth
stressing here that to obtain the above described relations we used the
evolution equations in order to eliminate the higher than the first time
derivatives of the metric perturbations in the RW gauge. 

An interesting application of these relations is to compute the
``close limit approximation" for black hole collisions by directly 
integrating the Teukolsky equation\cite{CKL98}. This constitutes  
a test for both, our relations on each hypersurface and 
the numerical accuracy of the code \cite{KLPA97} for integrating the 
Teukolsky equation. 

All this could be accomplished for the nonrotating case, which
provides the $a=0$ limit to be recovered by the more complicated 
(and realistic) case of $a\neq $ $0$, for which many steps should 
be generalized. In fact, in the rotating case 
we do not have the multipole decomposition 
of the metric perturbations, nor
a Regge-Wheeler gauge that can be reexpressed explicitly in terms of other
invariants of the 3-geometry and the extrinsic curvature only, neither the
generalization of $\phi _M$ for that case (that would be an
interesting result in itself). For these reasons a more geometrical
approach (making use of the Gauss-Codazzi relations) should be
followed in order to rewrite $\psi _4$ in the Kerr background\cite{CL98}.

A second aspect we have been able to study is the question of the
regularization of the Green function solution to the Teukolsky equation in
the {\it frequency }domain when nonvanishing initial data are present
(see Appendix). The
appropriate method to incorporate initial data in this case is to Laplace
transform the Teukolsky equation. In this way, initial data appear as an
additional source term and can be easily manipulated. We have explicitly
shown in the nonrotating case and using Brill-Lindquist initial data that $%
\ell =2$ and $\ell =3$ modes need to be regularized. This has to be a
general feature, valid also for the rotating background and for any other
kind of astrophysically reasonable initial data
for black hole collisions, and has to do with the fact
that realistic initial data are not of compact support (or do not decay fast
enough at spatial infinity).

\begin{acknowledgments}
C.O.L. is a member of the Carrera del Investigador Cient\'\i fico of
CONICET, Argentina and thanks FUNDACI\'ON ANTORCHAS for partial
financial support.
\end{acknowledgments}

\appendix

\section{REGULARIZATION OF THE INITIAL DATA}

In this appendix we shall consider again the problem of
divergent integrals that appear when in the frequency domain
one tries to solve the Teukolsky 
equation with sources that extend to infinity  
using the standard Green function technics. 
It was first proven by Poisson\cite{P97} that, for a source
generated by a particle released at rest from infinity (which has 
vanishing initial data), the divergent integrals appearing in the
formal solution can be regularized in the nonrotating case. 
Soon after, the present authors\cite{CL97} extended to the
rotating (Kerr) background Poisson's approach. 
Here, we show how the same problem appears when one consider 
bounded sources, but nonvanishing initial data, like Misner or 
Brill-Lindquist data\cite{M63,BL63} which are 
relevant to study black hole head-on collisions, in the close 
limit approximation. We thus apply the same regularization technics 
developed in references \cite{P97,CL97} to show that the Green function
formulation can be regularized.
We shall refer to the above two references for further details of the problem 
and computations, and below only sketch how things go for the case of 
nonvanishing initial data.

It is well known that the Teukolsky equation\ (\ref{master}) can be
separated by mode decomposition of the field and source into a complex
radial wave equation and an angular equation satisfied by the so-called
spin-weighted oblate spheroidal harmonics $_sS_\ell^m(\theta,a\omega)$ \cite
{PT73}, that in the case $a=0$ reduce\cite{GJN67} to the spin-weighted
spherical harmonics $%
_sY_\ell^m(\theta)$. By use of this separation of variables, the Teukolsky
equation, in its frequency domain, was widely studied in the early seventies%
\cite{T73,D79} providing extraordinary results like the proof of the
stability of rotating black hole and the computation of the gravitational
radiation for unbounded particle trajectories around rotating black holes.

A Fourier analysis in the complex plane can be used in the case one want to
study the problem of a particle falling into the black hole from any {\it %
finite} distance. In this case one must consider the initial value problem,
with $\Psi $ and $\partial_t \Psi $ specified on an initial
hypersurface (at $t=0)$.
To this purpose it is very useful to define the Laplace transform ${\bf %
\Psi }$ of $\Psi $ to be 
\begin{equation}
{\bf \Psi }(\omega ,r,\theta ,\varphi )\equiv \int_0^\infty e^{i\omega
t}\Psi (t,r,\theta ,\varphi )\,dt\ .  \label{Lform}
\end{equation}
We take $\Psi $ to vanish for $t<0$, which means that ${\bf \Psi }$ must be
analytic in the upper half of the complex $\omega $ plane.

Applying the Laplace transform to Eq.\ (\ref{master}), the Teukolsky
equation may be written in the following form 
\begin{eqnarray}
&&\ \ \ \Biggr\{\,\Delta ^{-s}\partial _r\left( \Delta ^{s+1}\partial
_r\right) -\omega ^2\left[ a^2\sin ^2\theta -\frac{(r^2+a^2)^2}\Delta
\right] +2is\omega \left[ (r+ia\cos \theta )-\frac{M(r^2-a^2)}\Delta \right]
\nonumber \\
&&\ \ \ +\frac 1{\sin \theta }\partial _\theta \left( \sin \theta \partial
_\theta \right) +\left[ \frac 1{\sin ^2\theta }-\frac{a^2}\Delta \right]
\partial _{\varphi \varphi }+2s\left[ \frac a\Delta (r-M+\frac{2i\omega Mr}s%
)+\frac{i\cos \theta }{\sin ^2\theta }\right] \partial _\varphi  \nonumber \\
\ &&-s\left( s\cot ^2\theta -1\right) \Biggr\}{\bf \Psi }= {\bf T}%
+S_{ID}~,  \label{Lteuk}
\end{eqnarray}
where 
\begin{eqnarray}
S_{ID} &\equiv &\left[ a^2\sin ^2\theta -\frac{(r^2+a^2)^2}\Delta \right]
\partial _t\Psi |_{t=0}-\Biggr\{i\omega \left[ a^2\sin ^2\theta -\frac{%
(r^2+a^2)^2}\Delta \right]  \nonumber \\
&&\ \ \ +4\frac{Mar}\Delta \partial _\varphi +2s\left[ (r+ia\cos \theta )-%
\frac{M(r^2-a^2)}\Delta \right] \Biggr\} \Psi |_{t=0}~.  \label{SID}
\end{eqnarray}
and 
\begin{equation}
{\bf T}(\omega ,r,\theta ,\varphi )\equiv \int_0^\infty e^{i\omega
t}4\pi \Sigma T(t,r,\theta ,\varphi )\,dt\ .  \label{LT}
\end{equation}
In this way Cauchy data are explicitly incorporated into the Teukolsky
equation as part of the source term. Eq.\ (\ref{Lteuk}) can be separated
into a radial and an angular part \cite{T72,CL97} and applied to study
particular problems involving finite radii infall. 

Let us consider the source term given by Eq. (\ref{SID}) in the nonrotating
case 
\begin{equation}
\frac{S_{\text{ID}}}{\Delta ^2}=\frac 1{r^4\left( 1-\frac{2M}r\right) ^3}%
\left[ \left( i\omega r^2+4r-12M\right) \psi _4(t=0,r)-r^2\partial _t\psi
_4(t=0,r)\right] \dot =\widetilde{G}g  \label{fuente1}
\end{equation}

where $\widetilde{G}=\sqrt{(\ell -1)\ell (\ell +1)(\ell +2)(2\ell +1)/(4\pi )%
}$ , and $g(r)$ takes the following form for the Brill-Lindquist initial data%
\cite{BL63}

\begin{equation}
g(r)=\frac{\left( i\omega r^2+4r-10M\right) }{\left( r-2M\right) ^2}\left( 
\frac{\stackrel{\_\_}{r}}r\right) ^{1/2}\sum_\ell \left( \frac M{\stackrel{%
\_\_}{r}}\right) ^{\ell +1}  \label{Gtilde}
\end{equation}
where $\stackrel{\_\_}{r}$ is the coordinate in the conformal space, that we
can take as the isotropic radial coordinate, related to $r$ as $\stackrel{%
\_\_}{r}=\left( \sqrt{r}+\sqrt{r-2M}\right) ^2/4$ .

This expression holds for the case of two equal masses black holes initially
at rest, and in the perturbative regime (close limit\cite{PP94,AP96b}), but
a very similar expression holds for the case of unequal masses (included the
particle limit). In particular, the conclusions about regularization will
hold the same also for different initial data such as the Misner ones\cite
{M63} (symmetrization through the throats).

As noted by Poisson\cite{P97} the divergent integral generated by the Green
method, $I_{{\rm div}}$ , can be regularized by introducing a function $h(r)$
and integrating by parts i. e. writing it in the following form 
\begin{equation}
I_{{\rm div}}=\int \biggl[\widetilde{G}g(r){\cal G}_{{\rm div}}e^{i\omega t}%
{\cal L}X^{H,\infty }+\frac d{dr}\left( he^{i\omega t}{\cal L}X^{H,\infty
}\right) \biggr]\,dr-he^{i\omega t}{\cal L}X^{H,\infty }\biggr|_{\text{%
boundaries}}  \label{R4}
\end{equation}
where ${\cal G}_{{\rm div}}=2i\omega r^2\left( 1-3M/r+i\omega r\right) $ , $%
{\cal L}=d/dr^{*}+i\omega $ (with $r^{*}=r+2M\ln \left( r/2M-1\right) $ ),
and $X^{H,\infty }$ are solutions to the Regge-Wheeler equation\cite{RW57}
(in the frequency domain) with purely ingoing and purely outgoing behavior
in the horizon and at infinity respectively. The function $h(r)$ is
determined by 
\[
h=-\widetilde{G}e^{-i\omega r^{*}}\int e^{i\omega r^{*}}g{\cal G}_{{\rm div}%
}dr. 
\]
By direct power counting in the first addend of Eq. (\ref{R4}) we observe
that the more important modes for computing gravitational radiation, modes $%
\ell =2$ and $\ell =3$ (in the unequal masses case) need to be regularized
since the integral diverges linearly and logarithmically, respectively.

\end{document}